\documentclass[11pt]{article}
\usepackage{amsmath,amssymb}

\newtheorem{theorem}{Theorem}
\newtheorem{lemma}{Lemma}
\newtheorem{proposition}{Proposition}

\begin{document}
\title{Symbolic transfer entropy rate is equal to transfer entropy rate for bivariate finite-alphabet stationary ergodic Markov processes}

\author{Taichi Haruna\footnote{Corresponding author}$\ ^{\rm ,1}$,  Kohei Nakajima$\ ^{\rm 2, 3}$ \\
\footnotesize{$\ ^{\rm 1}$ Department of Earth \& Planetary Sciences, Graduate School of Science, } \\
\footnotesize{Kobe University, 1-1, Rokkodaicho, Nada, Kobe 657-8501, Japan} \\
\footnotesize{$\ ^{\rm 2}$ Department of Informatics, } \\
\footnotesize{University of Zurich, Andreasstrasse 15, 8050 Zurich, Switzerland} \\
\footnotesize{$\ ^{\rm 3}$ Department of Mechanical and Process Engineering, } \\
\footnotesize{ETH Zurich, Leonhardstrasse 27, 8092 Zurich, Switzerland} \\
\footnotesize{E-mail: tharuna@penguin.kobe-u.ac.jp (T. Haruna)} 
}
\date{}

\maketitle

\abstract{
\textit{Transfer entropy} is a measure of the magnitude and the direction of information flow between 
jointly distributed stochastic processes. In recent years, its permutation analogues are considered 
in the literature to estimate the transfer entropy by counting the number of occurrences of orderings 
of values, not the values themselves. It has been suggested that the method of permutation 
is easy to implement, computationally low cost and robust to noise when applying to real world time series data. 
In this paper, we initiate a theoretical treatment of the corresponding rates. 
In particular, we consider the \textit{transfer entropy rate} and its permutation analogue, 
the \textit{symbolic transfer entropy rate}, and show that they are equal for any bivariate finite-alphabet 
stationary ergodic Markov process. 
This result is an illustration of the duality method introduced in [T. Haruna and K. Nakajima, Physica D 240, 1370 (2011)]. 
We also discuss the relationship among the transfer entropy rate, the time-delayed mutual information rate and their permutation analogues. 
}

\section{\label{sec1}Introduction}
Quantifying networks of information flows is critical to understand 
functions of complex systems such as living, social and technological systems. 
Schreiber \cite{Schreiber2000} introduced the notion of \textit{transfer entropy} to measure 
the magnitude and the direction of information flow from one element to 
another element emitting stationary signals in a given system. It has been 
used to analyze information flows in real time series data from neuroscience 
\cite{Schreiber2000,Besserve2010,Buehlmann2010,Garofalo2009,Honey2007,Katura2006,Luedtke2010,Lungarella2006,Neymotin2011,Sabesan2009,Vakorin2009}, 
and many other fields \cite{Bauer2007,Kamberaj2009,Marschinski2002,Materassi2007,Moniz2007a,Nichols2005,Pahle2008,Thai2007}.

The notion of \textit{permutation entropy} introduced by Bandt and Pompe \cite{Bandt2002a} has been 
proved that much of information contained in stationary time series can be captured by 
counting occurrences of orderings of values, not those of values themselves \cite{Bandt2002b,Bandt2007,Amigo2007,Misiurewicz2003,Keller2010,Keller2012}. 
The method of permutation has been applied across many disciplines \cite{Amigo2010} and 
suggested that it is easy to implement, computationally low cost and robust to noise when applying to real world time series data. 
Among the previous works, one relevant theoretical result to this paper is that the entropy rate \cite{Cover1991}, 
which is one of the most fundamental quantities of stationary stochastic processes, is equal to the permutation entropy rate for any finite-alphabet 
stationary stochastic process \cite{Amigo2005,Amigo2012}. 

The \textit{symbolic transfer entropy} \cite{Staniek2008} is a permutation analogue of the transfer entropy and 
has been used as an efficient and conceptually simple way of quantifying information flows 
in real time series data \cite{Staniek2008,Kowalski2010,Martini2011,Papana2011}. 
Another permutation analogue of the transfer entropy called \textit{transfer entropy on rank vectors} 
has been introduced to improve the performance of the symbolic transfer entropy \cite{Kugiumtzis2012}. 
So far, most of the work on permutation analogues of the transfer entropy are in application side. 
This paper concerns the theoretical relationship among respective rates. 
In particular, we consider the rate of transfer entropy on rank vectors which we call 
\textit{symbolic transfer entropy rate} and show that it is equal to 
the \textit {transfer entropy rate} \cite{Amblard2011} for any bivariate finite-alphabet stationary ergodic Markov process. 
We also discuss the relationship among the transfer entropy rate, the time-delayed mutual information rate 
and their permutation analogues. 

Our approach is based on the duality between values and orderings introduced by the authors \cite{Haruna2011}. 
In \cite{Haruna2011}, the excess entropy \cite{Crutchfield1983,Crutchfield2003,Feldman2008,Shaw1984,Grassberger1986,Bialek2001,Li1991,Arnold1996}, 
which is an effective measure of complexity of stationary stochastic processes, and its permutation analogue is shown to be equal 
for any finite-alphabet stationary ergodic Markov process. In this paper, we extend this approach to the bivariate case and address the relationship 
between the transfer entropy rate and the symbolic transfer entropy rate. 

This paper is organized as follows. 
In Section \ref{sec2}, we introduce the transfer entropy rate and the symbolic transfer entropy rate. 
We also discuss some combinatorial facts used in later sections. 
In Section \ref{sec3}, we give a proof of the equality between the transfer entropy rate and the 
symbolic transfer entropy rate which holds for bivariate finite-alphabet stationary ergodic Markov processes. 
In Section \ref{sec4}, we discuss the relationship among the transfer entropy rate, the 
time-delayed mutual information rate and their permutation analogues. 
Finally, in Section \ref{sec5}, we give concluding remarks. 

\section{\label{sec2}Definitions and Preliminaries}
Let $A_n=\{1,2,\cdots,n\}$ be a finite alphabet consisting of natural numbers from $1$ to $n$. 
In the following discussion, 
$\mathbf{X} \equiv \{X_1,X_2,\cdots \}$ and $\mathbf{Y} \equiv \{Y_1,Y_2,\cdots \}$ are jointly distributed 
finite-alphabet stationary stochastic processes, or equivalently, $(\mathbf{X},\mathbf{Y})$ is a bivariate finite-alphabet 
stationary stochastic process $\{(X_1,Y_1),(X_2,Y_2),\cdots \}$, 
where stochastic variables $X_i$ and $Y_j$ take their values in the alphabet $A_n$ and $A_m$, respectively. 
We use the notation $X_1^L \equiv (X_1,X_2,\cdots,X_L)$ for simplicity. 
We write $p(x_1^{L_1},y_1^{L_2})$ for the joint probability of the occurrence of words $x_1^{L_1} \equiv x_1 x_2 \cdots x_{L_1} \in A_n^{L_1}$ and 
$y_1^{L_2} \equiv y_1 y_2 \cdots y_{L_2} \in A_m^{L_2}$ for $L_1,L_2 \geq 1$. 

Originally, the notion of transfer entropy was introduced as a generalization of the entropy rate 
to bivariate processes \cite{Schreiber2000}. Along this original motivation, here, we do not consider 
the \textit{transfer entropy} but the \textit{transfer entropy rate} \cite{Amblard2011} from $\mathbf{Y}$ to $\mathbf{X}$ which is defined by 
\begin{equation}
t(\mathbf{X} | \mathbf{Y}) \equiv h(\mathbf{X}) - h(\mathbf{X} | \mathbf{Y}), 
\label{eq1}
\end{equation}
where $h(\mathbf{X}) \equiv \lim_{L \to \infty} H(X_1^L)/L$ is the entropy rate of $\mathbf{X}$, 
$H(X_1^L) \equiv - \sum_{x_1^L \in A_n^L} p(x_1^L) \log_2 p(x_1^L)$ is the Shannon entropy of 
the occurrences of words of length $L$ in $\mathbf{X}$ and $h(\mathbf{X} | \mathbf{Y})$ is 
the \textit{conditional entropy rate} of $\mathbf{X}$ given $\mathbf{Y}$ defined by 
\begin{equation}
h(\mathbf{X} | \mathbf{Y}) \equiv \lim_{L \to \infty} H(X_{L+1} | X_{1}^{L}, Y_{1}^{L}), 
\label{eq2}
\end{equation}
which always converges. $t(\mathbf{X} | \mathbf{Y})$ has the properties that 
(i) $0 \leq t(\mathbf{X} | \mathbf{Y}) \leq h(\mathbf{X})$ and (ii) 
$t(\mathbf{X} | \mathbf{Y})=0$ if $X_1^L$ is independent of $Y_1^L$ for all $L \geq 1$. 

In order to introduce the notion of \textit{symbolic transfer entropy rate}, 
we define a total order on the alphabet $A_n$ by the usual ``less-than-or-equal-to'' relationship. 
Let $\mathcal{S}_L$ be the set of all permutations of length $L \geq 1$. We consider each 
permutation $\pi$ of length $L$ as a bijection on the set $\{1,2,\cdots,L\}$. Thus, 
each permutation $\pi \in \mathcal{S}_L$ can be identified with the sequence $\pi(1)\cdots\pi(L)$. 
The {\it permutation type} $\pi \in \mathcal{S}_L$ of a given word $x_1^L \in A_n^L$ is defined 
by re-ordering $x_1,\cdots,x_L$ in ascending order, namely, $x_1^L$ \textit{is of type} $\pi$ if we have 
$x_{\pi(i)} \leq x_{\pi(i+1)}$ and $\pi(i) < \pi(i+1)$ when $x_{\pi(i)} = x_{\pi(i+1)}$ 
for $i =1,2,\cdots,L-1$. 
For example, $\pi(1)\pi(2)\pi(3)\pi(4)\pi(5)=41352$ for $x_1^5=24213 \in A_4^5$ because 
$x_4 x_1 x_3 x_5 x_2=12234$. The map $\phi_n : A_n^L \to \mathcal{S}_L$ sends each word $x_1^L$ 
to its unique permutation type $\pi=\phi_n(x_1^L)$. 

We will use the notions of \textit{rank sequences} and \textit{rank variables} \cite{Amigo2005}. 
The \textit{rank sequences} of length $L$ 
are words $r_1^L \in A_L^L$ satisfying $1 \leq r_i \leq i$ for $i=1,\cdots,L$. The set of 
all rank sequences of length $L$ is denoted by $\mathcal{R}_L$. It is clear that $|\mathcal{R}_L|=L!=|\mathcal{S}_L|$. 
Each word $x_1^L \in A_n^L$ can be mapped to a rank sequence $r_1^L$ by defining 
$r_i \equiv \sum_{j=1}^i \delta (x_j \leq x_i)$ for $i=1,\cdots,L$, where $\delta(P)=1$ if 
the proposition $P$ is true, otherwise $\delta(P)=0$. We denote this map from $A_n^L$ to 
$\mathcal{R}_L$ by $\varphi_n$. It can be shown that the map $\varphi_n : A_n^L \to \mathcal{R}_L$ 
is compatible with the map $\phi_n : A_n^L \to \mathcal{S}_L$ in the following sense: 
there exists a bijection $\iota : \mathcal{R}_L \to \mathcal{S}_L$ such that $\iota \circ \varphi_n=\phi_n$ \cite{Haruna2011}. 
The \textit{rank variables} associated with $\mathbf{X}$ are defined by 
$R_i \equiv \sum_{j=1}^i \delta (X_j \leq X_i)$ for $i=1,\cdots,L$. In general, $\mathbf{R} \equiv \{R_1,R_2,\cdots\}$ 
is a non-stationary stochastic process. 

The \textit{symbolic transfer entropy rate} from $\mathbf{Y}$ to $\mathbf{X}$ is defined by 
\begin{equation}
t^*(\mathbf{X} | \mathbf{Y}) \equiv h^*(\mathbf{X}) - h^*(\mathbf{X} | \mathbf{Y}), 
\label{eq3}
\end{equation}
where $h^*(\mathbf{X}) \equiv \lim_{L \to \infty} H^*(X_1^L)/L$ is the permutation entropy rate 
which is known to exist and is equal to $h(\mathbf{X})$ \cite{Amigo2005}, 
\begin{equation*}
H^*(X_1^L) \equiv - \sum_{\pi \in \mathcal{S}_L} p(\pi) \log_2 p(\pi) 
\end{equation*}
is the Shannon entropy of 
the occurrences of permutations of length $L$ in $\mathbf{X}$, $p(\pi)=\sum_{\phi_n(x_1^L)=\pi} p(x_1^L)$ 
and $h^*(\mathbf{X} | \mathbf{Y})$ is given by 
\begin{equation}
h^*(\mathbf{X} | \mathbf{Y}) \equiv \lim_{L \to \infty} \left( H^*(X_{1}^{L+1}, Y_{1}^{L}) - H^*(X_{1}^{L}, Y_{1}^{L}) \right) 
\label{eq4}
\end{equation}
if the limit in the right hand side exists. 
Here, $H^*(X_{1}^{L_1}, Y_{1}^{L_2})$ is defined by 
\begin{equation*}
H^*(X_{1}^{L_1}, Y_{1}^{L_2}) \equiv - \sum_{\pi \in \mathcal{S}_{L_1}, \pi' \in \mathcal{S}_{L_2}} p(\pi,\pi') \log_2 p(\pi,\pi'), 
\end{equation*}
where $p(\pi,\pi')=\sum_{\phi_n(x_1^{L_1})=\pi,\phi_m(y_1^{L_2})=\pi'} p(x_1^{L_1},y_1^{L_2})$. 

Let $\mathbf{R}$ and $\mathbf{S}$ be rank variables associated with $\mathbf{X}$ and $\mathbf{Y}$, respectively. 
By the compatibility between $\phi_k$ and $\varphi_k$ for $k=m,n$, we have $H(R_1^{L_1},S_1^{L_2})=H^*(X_1^{L_1},Y_1^{L_2})$. 
Thus, $h^*(\mathbf{X} | \mathbf{Y})$ can be written as 
\begin{equation*}
h^*(\mathbf{X} | \mathbf{Y})=\lim_{L \to \infty} H(R_{L+1} | R_1^L,S_1^L) 
\end{equation*}
if $h^*(\mathbf{X} | \mathbf{Y})$ exists. 

Note that the above definition of the symbolic transfer entropy rate (\ref{eq3}) is not the rate of 
the original symbolic transfer entropy introduced by Staniek and Lehnertz \cite{Staniek2008} but that of 
the transfer entropy on rank vectors \cite{Kugiumtzis2012} which is an improved version of it. 

\section{\label{sec3}Main Result}
In this section, we give a proof of the following theorem: 

\begin{theorem}
For any bivariate finite-alphabet stationary ergodic Markov process $(\mathbf{X},\mathbf{Y})$, 
we have the equality 
\begin{equation}
t(\mathbf{X} | \mathbf{Y}) = t^*(\mathbf{X} | \mathbf{Y}). 
\label{eq5}
\end{equation}
\label{thm1}
\end{theorem}

Before proceeding to the proof of Theorem \ref{thm1}, first 
we present some intermediate results used in the proof. 

We introduce the map $\mu : \mathcal{S}_L \to \mathbb{N}^L$, where $\mathbb{N}=\{1,2,\cdots\}$ is 
the set of all natural numbers ordered by usual ``less-than-or-equal-to'' relationship, 
by the following procedure: first, given a permutation $\pi \in \mathcal{S}_L$, 
we decompose the sequence $\pi(1) \cdots \pi(L)$ into \textit{maximal ascending subsequences}. 
A subsequence $i_j \cdots i_{j+k}$ of a sequence $i_1 \cdots i_L$ 
is called a \textit{maximal ascending subsequence} if it is ascending, namely, 
$i_j \leq i_{j+1} \leq \cdots \leq i_{j+k}$, and neither $i_{j-1} i_{j} \cdots i_{j+k}$ nor 
$i_{j} i_{j+1} \cdots i_{j+k+1}$ is ascending. 
Second, if $\pi(1) \cdots \pi(i_1), \ \pi(i_1+1) \cdots \pi(i_2), \cdots, \pi(i_{k-1}+1) \cdots \pi(L)$ 
is the decomposition of $\pi(1)\cdots\pi(L)$ into maximal ascending subsequences, then we define 
the word $x_1^L \in \mathbb{N}^L$ by 
$x_{\pi(1)}=\cdots=x_{\pi(i_1)}=1, x_{\pi(i_1+1)}=\cdots=x_{\pi(i_2)}=2, \cdots, x_{\pi(i_{k-1}+1)}=\cdots=x_{\pi(L)}=k$. 
Finally, we define $\mu(\pi)=x_1^L$. 
For example, the decomposition of $25341 \in \mathcal{S}_5$ into maximal ascending subsequences is 
$25,34,1$. We obtain $\mu(\pi)=x_1 x_2 x_3 x_4 x_5=31221$ by putting $x_2 x_5 x_3 x_4 x_1=11223$. 
By construction, we have $\phi_n \circ \mu(\pi)=\pi$ when $\mu(\pi) \in A_n^L$ 
for any $\pi \in \mathcal{S}_L$. 

The map $\mu$ can be seen as the dual to the map $\phi_n$ (or $\varphi_n$) in the following sense: 
\begin{theorem}[Theorem 9 in \cite{Haruna2011}] 
Let us put 
\begin{eqnarray*}
B_{n,L} &\equiv& \{x_1^L \in A_n^L | \exists \pi \in \mathcal{S}_L \textrm{ such that } \phi_n^{-1}(\pi)=\{x_1^L\} \}, \\
C_{n,L} &\equiv& \{\pi \in \mathcal{S}_L | |\phi_n^{-1}(\pi)|=1 \}, 
\end{eqnarray*}
where 
$\phi_n^{-1}(\pi) \equiv \{ x_1^L \in A_n^L | \phi_n(x_1^L)=\pi \}$ is the inverse image of $\pi \in \mathcal{S}_L$ 
by the map $\phi_n$. 
Then, 
\begin{itemize}
\item[(i)]
$\phi_n$ restricted on $B_{n,L}$ is a map into $C_{n,L}$, $\mu$ restricted on $C_{n,L}$ is a map into $B_{n,L}$, 
and they form a pair of mutually inverse maps. 
\item[(ii)] $x_1^L \in B_{n,L}$ if and only if 
\begin{eqnarray}
\textrm{for all } 1 \leq i \leq n-1 \textrm{ there exist } 1 \leq j < k \leq L \nonumber \\
\textrm{ such that } x_j=i+1 \textrm{ and } x_k=i. 
\label{eq6}
\end{eqnarray}
\end{itemize}
\label{thm2}
\end{theorem}

The proof of Theorem \ref{thm2} can be found in \cite{Haruna2011}. 

Since $h(\mathbf{X})=h^*(\mathbf{X})$ holds for any finite-alphabet stationary process, 
proving (\ref{eq5}) is equivalent to showing that the equality 
\begin{equation}
\lim_{L \to \infty} H(R_{L+1} | R_1^L,S_1^L) = \lim_{L \to \infty} H(X_{L+1} | X_1^L,Y_1^L) 
\label{eq7}
\end{equation}
holds for any bivariate finite-alphabet stationary ergodic Markov process $(\mathbf{X},\mathbf{Y})$. 
For simplicity, we assume that each $(x,y) \in A_n \times A_m$ appears with 
a positive probability $p(x,y)>0$. The essentially same proof can be applied to the general case. 

\begin{lemma}
For any $\epsilon >0$ if we take $L$ sufficiently large, then 
\begin{eqnarray}
\sum_{\begin{subarray}{c} x_1^L \textrm{ satisfies } (*),\\ y_1^L \textrm{ satisfies }(**) \end{subarray}} p(x_1^L,y_1^L) > 1-\epsilon, 
\label{eq8}
\end{eqnarray}
where $(*)$ is the condition that for any $x \in A_n$ there exist 
$1 \leq i \leq \lfloor L/2 \rfloor < j \leq L$ such that $x=x_i=x_j$ and 
$(**)$ is the condition that for any $y \in A_m$ there exist 
$1 \leq i' \leq \lfloor L/2 \rfloor < j' \leq L$ such that $y=y_{i'}=y_{j'}$. 
\label{lem1}
\end{lemma}

\textit{Proof. }
The ergodicity of $(\mathbf{X},\mathbf{Y})$ implies that the relative frequency of any word 
$(x_1^k,y_1^k)$ converges in probability to $p(x_1^k,y_1^k)$. 
In particular, if $F_{(x,y)}^N$ is the stochastic variable defined by the number of indexes $1 \leq i \leq N$ 
such that $(X_i,Y_i)=(x,y)$ for $(x,y) \in A_n \times A_m$, then we have 
for any $\epsilon>0$ and $\delta>0$ there exists $N_{(x,y),\epsilon,\delta}$ such that if 
$N > N_{(x,y),\epsilon,\delta}$ then 
\begin{equation*}
\mathrm{Pr}\{|F_{(x,y)}^N/N - p(x,y)| < \delta\} > 1-\epsilon. 
\end{equation*}

Now, fix any $\epsilon>0$. Choose $\delta$ so that 
\begin{equation*}
0 < \delta < \min_{(x,y) \in A_n \times A_m}\{p(x,y)\} 
\end{equation*}
and 
put $N_0 \equiv \max_{(x,y) \in A_n \times A_m} \{ N_{(x,y),\epsilon/(2nm),\delta} \}$. 
Let $S_{(x,y)}^N$ be the set of words $(x_1^N,y_1^N)$ such that there exists $1 \leq i \leq N$ that satisfies 
$x_i=x$ and $y_i=y$, and $S_N$ the set of words $(x_1^N,y_1^N)$ such that 
for any $(x,y) \in A_n \times A_m$ there exists $1 \leq i \leq N$ that satisfies $x_i=x$ and $y_i=y$. 

If $N>N_0$, then we have for any $(x,y) \in A_n \times A_m$ 
\begin{align*}
\mathrm{Pr}(S_{(x,y)}^N) &= \sum_{(x_1^N,y_1^N) \in S_{(x,y)}^N} p(x_1^N,y_1^N) \\
&= \mathrm{Pr}\{ F_{(x,y)}^N >0 \} \\
&\geq \mathrm{Pr}\{|F_{(x,y)}^N/N - p(x,y)| < \delta\} \\
&> 1-\epsilon/(2nm), 
\end{align*}
where the inequality in the third line holds follows because we have 
$p(x,y) > \delta$ by the choice of $\delta$. 

Then, having that 
\begin{equation*}
S_N \equiv \bigcap_{(x,y) \in A_n \times A_m} S_{(x,y)}^N, 
\end{equation*}
it follows that 
\begin{equation*}
\mathrm{Pr}(S_N) > 1 - nm \times \epsilon/(2nm)=1-\epsilon/2. 
\end{equation*}

Now, take $L$ so that $\lfloor L/2 \rfloor > N_0$. Let $U$ be the set of words $(x_1^L,y_1^L)$ 
such that $(x_1^{\lfloor L/2 \rfloor},y_1^{\lfloor L/2 \rfloor}) \in S_{\lfloor L/2 \rfloor}$ and 
$V$ the set of words $(x_1^L,y_1^L)$ 
such that $(x_{\lfloor L/2 \rfloor + 1}^L,y_{\lfloor L/2 \rfloor +1}^L) \in S_{L - \lfloor L/2 \rfloor}$. 
Then, we have 
\begin{equation*}
\mathrm{Pr}(U) \geq \mathrm{Pr}(S_{\lfloor L/2 \rfloor})>1-\epsilon/2 
\end{equation*}
and 
\begin{equation*}
\mathrm{Pr}(V) \geq \mathrm{Pr}(S_{L-\lfloor L/2 \rfloor})>1-\epsilon/2. 
\end{equation*}
Consequently, we obtain 
\begin{equation*}
\sum_{\begin{subarray}{c} x_1^L \textrm{ satisfies } (*),\\ y_1^L \textrm{ satisfies }(**) \end{subarray}} p(x_1^L,y_1^L) 
= \mathrm{Pr}(U \cap V) 
> 1- \epsilon. 
\end{equation*}

\hfill $\Box$\\

We put 
\begin{eqnarray*}
D_{n,m,L} \equiv \{ (x_1^L,y_1^L) | x_1^L \textrm{ satisfies } (*) \textrm{ and } y_1^L \textrm{ satisfies } (**) \}
\end{eqnarray*}
and 
\begin{eqnarray*}
E_{n,m,L} \equiv \{ (r_1^L,s_1^L) | \exists (x_1^L,y_1^L) \in D_{n,m,L} \textrm{ such that } \\ \varphi_n(x_1^L)=r_1^L, \ \varphi_m(y_1^L)=s_1^L \}. 
\end{eqnarray*}
Then, we have $x_1^L \in B_{n,L}$ and $y_1^L \in B_{m,L}$ for any $(x_1^L,y_1^L) \in D_{n,m,L}$. 
Indeed, if $(x_1^L,y_1^L) \in D_{n,m,L}$, then $x_1^L$ and $y_1^L$ satisfy $(*)$ and $(**)$, respectively. 
For any $1 \leq i \leq n-1$, there exists $1 \leq j \leq \lfloor L/2 \rfloor$ such that $x_j=i+1$ and 
there exists $\lfloor L/2 \rfloor < k \leq L$ such that $x_k=i$ by $(*)$. Hence, $x_1^L$ satisfies (\ref{eq6}). 
By Theorem \ref{thm2} (ii), we have $x_1^L \in B_{n,L}$. By the same way, we have $y_1^L \in B_{m,L}$. 

Thus, the map 
\begin{equation*}
(x_1^L,y_1^L) \mapsto (\varphi_n(x_1^L),\varphi_m(y_1^L)) 
\end{equation*}
is a bijection from $D_{n,m,L}$ to $E_{n,m,L}$ due to 
the duality between $\phi_k$ and $\mu$ for $k=m,n$. 
Indeed, it is onto because 
$E_{n,m,L}$ is the image of the map $\varphi_n \times \varphi_m:A_n^L \times A_m^L \to \mathcal{R}_L \times \mathcal{R}_L$ 
restricted on $D_{n,m,L}$. It is also injective. For if 
$(\varphi_n(x_1^L),\varphi_m(y_1^L))=(\varphi_n(\overline{x}_1^L),\varphi_m(\overline{y}_1^L))$, then 
$\phi_n(x_1^L)=\iota \circ \varphi_n(x_1^L)=\iota \circ \varphi_n(\overline{x}_1^L)=\phi_n(\overline{x}_1^L)$ and similarly 
$\phi_m(y_1^L)=\phi_m(\overline{y}_1^L)$. By Theorem \ref{thm2} (i), $\phi_n$ and $\phi_m$ are bijections from $B_{n,L}$ to $C_{n,L}$ and 
from $B_{m,L}$ to $C_{m,L}$, respectively. Since $x_1^L,\overline{x}_1^L \in B_{n,L}$ and $y_1^L,\overline{y}_1^L \in B_{m,L}$, 
it hold that $x_1^L=\overline{x}_1^L$ and $y_1^L=\overline{y}_1^L$. 

In particular, we have 
\begin{equation*}
p(x_1^L,y_1^L)=p(r_1^L,s_1^L) 
\end{equation*}
and 
\begin{equation*}
p(r_{L+1} | r_1^L,s_1^L)=p(r_{L+1} | x_1^L,y_1^L) 
\end{equation*}
for any $(x_1^L,y_1^L) \in D_{n,m,L}$, 
where $r_1^L=\varphi_n(x_1^L)$ and $s_1^L=\varphi_m(y_1^L)$. 

\textit{Proof of Theorem \ref{thm1}.} 
Given any $\epsilon > 0$, let us take $L$ large enough so that the inequality (\ref{eq8}) holds. 
We shall evaluate each term in the right hand side of (\ref{eq9}): 
\begin{multline}
H(X_{L+1} | X_1^L, Y_1^L) - H(R_{L+1} | R_1^L, S_1^L) \\
= - \sum_{(x_1^L,y_1^L) \in D_{n,m,L}} p(x_1^L,y_1^L) 
\Biggl( \sum_{x_{L+1}} p(x_{L+1} | x_1^L,y_1^L) \log_2 p(x_{L+1} | x_1^L,y_1^L) \\
- \sum_{r_{L+1}} p(r_{L+1} | x_1^L,y_1^L) \log_2 p(r_{L+1} | x_1^L,y_1^L) \Biggr) \\
- \sum_{(x_1^L,y_1^L) \not \in D_{n,m,L}} p(x_1^L,y_1^L) \sum_{x_{L+1}} p(x_{L+1} | x_1^L,y_1^L) \log_2 p(x_{L+1} | x_1^L,y_1^L) \\
+ \sum_{(r_1^L,s_1^L) \not \in E_{n,m,L}} p(r_1^L,s_1^L) \sum_{r_{L+1}} p(r_{L+1} | r_1^L,s_1^L) \log_2 p(r_{L+1} | r_1^L,s_1^L). 
\label{eq9}
\end{multline}
First, the second term in (\ref{eq9}) is bounded by $\epsilon \log_2 n$ which can be arbitrary small. 
This is because 
\begin{equation*}
\sum_{(x_1^L,y_1^L) \not \in D_{n,m,L}} p(x_1^L,y_1^L) \leq \epsilon 
\end{equation*}
by Lemma \ref{lem1} and the sum over $x_{L+1}$ is at most $\log_2 n$. 

Second, to show the third term also converges to $0$ as $L \to \infty$, we use the Markov property: 
if $(\mathbf{X},\mathbf{Y})$ is ergodic Markov, then we can show that 
\begin{align*}
\sum_{(r_1^L,s_1^L) \not \in E_{n,m,L}} p(r_1^L,s_1^L)
&= \sum_{(x_1^L,y_1^L) \not \in D_{n,m,L}} p(x_1^L,y_1^L) \\
&< C \lambda^L 
\end{align*}
for some $C>0$ and $0 \leq \lambda <1$. 
Indeed, we have 
\begin{align*}
\sum_{(x_1^L,y_1^L) \not \in D_{n,m,L}} p(x_1^L,y_1^L) 
\leq \sum_{\begin{subarray}{c} x_1^L \textrm{ does not}\\ \textrm{satisfy } (*) \end{subarray}} p(x_1^L) 
+ \sum_{\begin{subarray}{c} y_1^L \textrm{ does not}\\ \textrm{satisfy } (**) \end{subarray}} p(y_1^L). 
\end{align*}
Since 
\begin{align*}
\sum_{\begin{subarray}{c} x_1^L \textrm{ does not}\\ \textrm{satisfy } (*) \end{subarray}} p(x_1^L) 
&\leq \sum_{x \in A_n} \left( \sum_{\begin{subarray}{c} x_i \neq x, \\ 1 \leq i \leq N \end{subarray}} p(x_1^N) + 
\sum_{\begin{subarray}{c} x_i \neq x, \\ N < i \leq L \end{subarray}} p(x_{N+1}^L) \right) \\
&\leq 2 \sum_{x \in A_n} \sum_{\begin{subarray}{c} x_i \neq x, \\ 1 \leq i \leq N \end{subarray}} p(x_1^N)
\end{align*} 
and similarly 
\begin{align*}
\sum_{\begin{subarray}{c} y_1^L \textrm{ does not}\\ \textrm{satisfy } (**) \end{subarray}} p(y_1^L) 
\leq 2 \sum_{y \in A_m} \sum_{\begin{subarray}{c} y_i \neq y, \\ 1 \leq i \leq N \end{subarray}} p(y_1^N), 
\end{align*} 
where $N=\lfloor L/2 \rfloor$, it is sufficient to show that the probabilities 
\begin{eqnarray*}
\beta_{x,\mathbf{X},L} \equiv \sum_{\begin{subarray}{c} x_i \neq x, \\ 1 \leq i \leq N \end{subarray} } p(x_1^N) 
\end{eqnarray*}
for all $x \in A_n$ and 
\begin{eqnarray*}
\beta_{y,\mathbf{Y},L} \equiv \sum_{\begin{subarray}{c} y_i \neq y, \\ 1 \leq i \leq N \end{subarray} } p(y_1^N) 
\end{eqnarray*}
for all $y \in A_m$ converge to $0$ exponentially fast. 

Let $P$ be the transition matrix for the Markov process $(\mathbf{X},\mathbf{Y})$. We denote 
its $(x,y)(x',y')$-th element by $p_{(x,y)(x',y')}$ which indicates the transition probability 
from state $(x,y)$ to $(x',y')$. We denote the stationary distribution associated with $(\mathbf{X},\mathbf{Y})$ 
by $\mathbf{p}=(p_{(x,y)})_{(x,y) \in A_n \times A_m}$ which uniquely exists because of the 
ergodicity of the process. The probability of the occurrence of a word $(x_1^L,y_1^L)$ is given by 
$p(x_1^L,y_1^L)=p_{(x_1,y_1)}p_{(x_1,y_1)(x_2,y_2)} \cdots p_{(x_{L-1},y_{L-1})(x_L,y_L)}$. 
For any $x \in A_n$, we define the 
matrix $P_x$ whose $(x',y')(x'',y'')$-th element is defined by 
\begin{eqnarray*}
(P_x)_{(x',y')(x'',y'')}=
\begin{cases}
0 & \textrm{if } x'=x \\
p_{(x',y')(x'',y'')} & \textrm{otherwise.}
\end{cases}
\end{eqnarray*}
Then, we can write 
\begin{eqnarray*}
\beta_{x,\mathbf{X},L}=\langle (P_x)^{N-1}\mathbf{u}_x, \mathbf{p} \rangle, 
\end{eqnarray*}
where the vector $\mathbf{u}_x=(u_{(x',y')})$ is defined by $u_{(x',y')}=0$ if $x'=x$ and 
otherwise $u_{(x',y')}=1$ and $\langle \cdots \rangle$ is the usual inner product in the 
$n \times m$-dimensional Euclidean space. 
Since $P_x$ is a non-negative matrix, we can apply the Perron-Frobenius theorem to it. 
We can show that its Perron-Frobenius eigenvalue (the non-negative eigenvalue whose absolute value 
is the largest among the eigenvalues) $\lambda_x$ is strictly less than $1$ by the same manner 
as in the proof of Lemma 13 in \cite{Haruna2011}. 
We can also show that for any $\delta>0$ there exists $C_{\delta,x}>0$ such that for any $k \geq 1$ 
\begin{eqnarray*}
\| (P_x)^k \mathbf{u}_x \| \leq C_{\delta,x} (\lambda_x + \delta)^k \| \mathbf{u}_x \|, 
\end{eqnarray*}
where $\| \cdots \|$ is the Euclidean norm. The proof for this fact is found in, 
for example, Section 1.2 of \cite{Katok1995}. 
Hence, if we choose $\delta>0$ sufficiently small 
so that $\lambda_x + \delta <1$ and put $\gamma_x \equiv (\lambda_x + \delta)^{1/2}$ and 
$C_x=C_{\delta,x}(\lambda_x + \delta)^{-2} \| \mathbf{u}_x \| \| \mathbf{p} \|$, 
then we have $\beta_{x,\mathbf{X},L} \leq C_x \gamma_x^L$. 
By the same manner, we can obtain the similar bound for $\beta_{y,\mathbf{Y},L}$ for all $y \in A_m$. 

Since the sum over $r_{L+1}$ is at most $\log_2 (L+1)$, 
the absolute value of the third term is bounded by the quantity $C \lambda^L \log_2 (L+1)$ which goes to $0$ as $L \to \infty$. 
Note that there is a $O(\log L)$ diverging term coming from the sum over $r_{L+1}$. The assumed ergodic Markov property is used to 
overcome this divergence by showing the quantity 
\begin{equation*}
\sum_{(r_1^L,s_1^L) \not \in E_{n,m,L}} p(r_1^L,s_1^L)
\end{equation*}
converges to 0 exponentially fast. 

Finally, the first term is shown to be $0$ by the same discussion as in the proof of Lemma 1 in \cite{Amigo2005} 
: if $(x_1^L,y_1^L) \in D_{n,m,L}$, then each symbol $x \in A_n$ 
appears at least once in the word $x_1^L$ (indeed, it appears at least twice). If $a_x$ is the number of 
$1 \leq x \leq n$ occurring in the word $x_1^L$, then $a_x>0$ for all $1 \leq x \leq n$. Hence, 
given $(x_1^L,y_1^L) \in D_{n,m,L}$, $x_{L+1}=x$ if and only if $r_{L+1}=1+\sum_{x'=1}^x a_{x'}$. 
Indeed, we have 
\begin{equation*}
r_{L+1} = \sum_{i=1}^{L+1} \delta(x_i \leq x_{L+1}) = 1 + \sum_{x'=1}^{x_{L+1}} a_{x'}. 
\end{equation*}
Hence, if $x_{L+1}=x$, then we have $r_{L+1} = 1 + \sum_{x'=1}^x a_{x'}$. 
For the converse, if $r_{L+1}=1+\sum_{x'=1}^x a_{x'}$, 
then we have $\sum_{x'=1}^{x_{L+1}} a_{x'}=\sum_{x'=1}^x a_{x'}$. 
Since $a_{x'}>0$ for all $1 \leq x' \leq n$, this happens only when $x_{L+1}=x$. 

Thus, given $(x_1^L,y_1^L) \in D_{n,m,L}$, the probability distribution 
\begin{equation*}
p(r_{L+1}|x_1^L,y_1^L) 
\end{equation*}
is just a re-indexing of $p(x_{L+1}|x_1^L,y_1^L)$, which implies that the first term is exactly equal to $0$. 
This completes the proof of the theorem. 

\hfill $\Box$\\

From the proof, we can also see that 
$t^*(\mathbf{X} | \mathbf{Y}) \leq t(\mathbf{X} | \mathbf{Y})$ holds for any bivariate finite-alphabet stationary ergodic process 
$(\mathbf{X}, \mathbf{Y})$ if $h^*(\mathbf{X} | \mathbf{Y})$ exists for the process. 

\section{\label{sec4}On the relationship with the time-delayed mutual information rate}
Apart from permutation, it is natural to ask whether the equality for the conditional entropy rate 
\begin{equation}
\lim_{L \to \infty} H(X_{L+1} | X_{1}^{L}, Y_{1}^{L}) = \lim_{L \to \infty} \frac{1}{L} H(X_1^{L+1} | Y_1^L) 
\label{eq10}
\end{equation}
holds or not, which is parallel to the equality for the entropy rate 
$\lim_{L \to \infty} H(X_{L+1} | X_{1}^{L}) = \lim_{L \to \infty} \frac{1}{L} H(X_1^{L+1})$
which holds for any finite-alphabet stationary stochastic process $\mathbf{X}$ \cite{Cover1991}. 
In this section, we will see that this question has an intimate relationship with the relationship between 
the transfer entropy rate and the \textit{time-delayed mutual information rate}.

In general, (\ref{eq10}) does not hold. For example, if $\mathbf{X}=\mathbf{Y}$, then we have 
$\lim_{L \to \infty} H(X_{L+1} | X_{1}^{L}, Y_{1}^{L})=h(\mathbf{X})$, 
while $\lim_{L \to \infty} \frac{1}{L} H(X_1^{L+1} | Y_1^L)=0$. 
However, note that the inequality 
\begin{equation}
\lim_{L \to \infty} H(X_{L+1} | X_{1}^{L}, Y_{1}^{L}) \geq \lim_{L \to \infty} \frac{1}{L} H(X_1^{L+1} | Y_1^L) 
\label{eq10-1}
\end{equation}
holds for any bivariate finite-alphabet stationary stochastic process $(\mathbf{X},\mathbf{Y})$. 
Indeed, we have 
\begin{eqnarray*}
&&\lim_{L \to \infty} H(X_{L+1} | X_{1}^{L}, Y_{1}^{L}) \\
&=&\lim_{L \to \infty} \frac{1}{L} \sum_{i=1}^{L+1} H(X_{i} | X_1^{i-1},Y_1^{i-1}) \\
&\geq& \lim_{L \to \infty} \frac{1}{L} \sum_{i=1}^{L+1} H(X_{i} | X_1^{i-1},Y_1^L) \\
&=&\lim_{L \to \infty} \frac{1}{L} H(X_1^{L+1} | Y_1^L), 
\end{eqnarray*}
where the first equality is due to the Ces\'aro mean theorem 
(if $\lim_{L \to \infty} b_L=b$ then $\lim_{L \to \infty} \frac{1}{L} \sum_{i=1}^L b_i=b$) 
and the last equality follows from the chain rule for the Shannon entropy. 
In the following, we give a sufficient condition for (\ref{eq10}). 

\begin{proposition}
If there exists $N>0$ such that if $i>N$ then $X_i$ is independent of $Y_i^{i+j}$ given $X_1^{i-1}$ and $Y_1^{i-1}$ for any $j \geq 0$, that is, 
\begin{align*}
& \mathrm{Pr}(X_i=x_i,Y_i^{i+j}=y_i^{i+j} | X_1^{i-1}=x_1^{i-1},Y_1^{i-1}=y_1^{i-1}) \\
&= \mathrm{Pr}(X_i=x_i | X_1^{i-1}=x_1^{i-1},Y_1^{i-1}=y_1^{i-1}) \\
&\times \mathrm{Pr}(Y_i^{i+j}=y_i^{i+j} | X_1^{i-1}=x_1^{i-1},Y_1^{i-1}=y_1^{i-1})
\end{align*} 
for any $j \geq 0$, 
$x_k \in A_n \ (1 \leq k \leq i)$ and $y_l \in A_m \ (1 \leq l \leq i+j)$, then (\ref{eq10}) holds, namely, we have 
the equality 
\begin{equation*}
\lim_{L \to \infty} H(X_{L+1} | X_{1}^{L}, Y_{1}^{L}) = \lim_{L \to \infty} \frac{1}{L} H(X_1^{L+1} | Y_1^L). 
\end{equation*}
\label{pro1}
\end{proposition}
\textit{Proof. }
Let us put $a_{i,L} \equiv H(X_{i+1}|X_1^{i},Y_1^L)$. If we fix the index $i$, then $a_{i,L}$ is a decreasing sequence of $L$. 
By the chain rule for the Shannon entropy, we have 
\begin{equation*}
H(X_1^{L+1}|Y_1^L) = 
\sum_{i=0}^L H(X_{i+1}|X_1^i,Y_1^L) = \sum_{i=0}^{L} a_{i,L}. 
\end{equation*}
However, by the assumption, we have $a_{i,i}=a_{i,i+1}=a_{i,i+2}=\cdots$ for $i>N$. 
Hence, we have 
\begin{equation*}
H(X_1^{L+1}|Y_1^L)=\sum_{i=0}^N a_{i,L} + \sum_{i=N+1}^L a_{i,i}. 
\end{equation*}
Since the former sum is finite, by the Ces\'aro mean theorem, 
we obtain 
\begin{align*}
\lim_{L \to \infty} \frac{1}{L} H(X_1^{L+1} | Y_1^L) 
&= \lim_{L \to \infty} \frac{1}{L} \sum_{i=N+1}^L a_{i,i} \\
&= \lim_{L \to \infty} a_{L,L} \\
&= \lim_{L \to \infty} H(X_{L+1} | X_{1}^{L}, Y_{1}^{L}). 
\end{align*}

\hfill $\Box$\\

Note that if the assumption holds, then it holds for $N=1$ by stationarity. 
If $(\mathbf{X},\mathbf{Y})$ is a stationary Markov process, then we can show by direct calculation that 
the assumption of Proposition \ref{pro1} is equivalent to the following simpler condition 
by using the Markov property: 
\begin{equation}
p(x_2,y_2 | x_1,y_1)=p(x_2 | x_1,y_1)p(y_2 | x_1,y_1) 
\label{eq11}
\end{equation}
for any $x_1,x_2 \in A_n$ and $y_1,y_2 \in A_m$. 

If (\ref{eq10}) holds, then we obtain 
\begin{equation}
t(\mathbf{X} | \mathbf{Y})= \lim_{L \to \infty} \frac{1}{L}I(X_1^{L+1};Y_1^L), 
\label{eq12}
\end{equation}
where $I(\mathbf{A};\mathbf{B})$ is the mutual information between stochastic variables $\mathbf{A}$ and $\mathbf{B}$. 
We call the quantity at the right hand side of (\ref{eq12}) 
\textit{time-delayed mutual information rate} and denote it by $i_{+1}(\mathbf{X};\mathbf{Y})$. 
Note that we have 
\begin{equation*}
t(\mathbf{X} | \mathbf{Y}) \leq i_{+1}(\mathbf{X};\mathbf{Y}) 
\end{equation*}
for any bivariate finite-alphabet stationary stochastic process $(\mathbf{X},\mathbf{Y})$ 
by the inequality (\ref{eq10-1}). 

For any bivariate finite-alphabet stationary stochastic process $(\mathbf{X},\mathbf{Y})$, 
it is straightforward to see that 
\begin{equation*}
i_{+1}(\mathbf{X} ; \mathbf{Y}) = h(\mathbf{X}) + h(\mathbf{Y}) - h(\mathbf{X},\mathbf{Y}) 
\end{equation*}
holds. Hence, we have 
\begin{equation*}
i_{+1}(\mathbf{X} ; \mathbf{Y}) = i_{+1}(\mathbf{Y} ; \mathbf{X}) = i(\mathbf{X} ; \mathbf{Y}) = i(\mathbf{Y} ; \mathbf{X}). 
\end{equation*}
Here, 
\begin{equation*}
i(\mathbf{X} ; \mathbf{Y}) \equiv \lim_{L \to \infty} \frac{1}{L}I(X_1^L;Y_1^L) 
\end{equation*}
is the \textit{mutual information rate} between $\mathbf{X}$ and $\mathbf{Y}$. 
Thus, when we consider the \textit{rate} for mutual information 
between two jointly distributed finite-alphabet stationary stochastic processes $\mathbf{X}$ and $\mathbf{Y}$, 
which is defined in the limit $L \to \infty$, time delay has no significance in contrast to 
the time-delayed mutual information which has been used in time series embedding \cite{Fraser1986} and 
detection of nonlinear interdependence between two time series at different time points \cite{Kaneko1986,Vastano1988}. 

The result on the relationship between 
$t(\mathbf{X} | \mathbf{Y})$ and $i_{+1}(\mathbf{X};\mathbf{Y})$ 
when $(\mathbf{X},\mathbf{Y})$ is a stationary Markov process can be summarized as follows: 

\begin{proposition}
Let $(\mathbf{X},\mathbf{Y})$ be a bivariate finite-alphabet stationary ergodic Markov process over the 
alphabet $A_n \times A_m$. 
If the condition 
\begin{equation*}
p(x_2,y_2 | x_1,y_1)=p(x_2 | x_1,y_1)p(y_2 | x_1,y_1) 
\end{equation*}
holds for any $x_1,x_2 \in A_n$ and $y_1,y_2 \in A_m$, 
then we have the equality 
\begin{equation*}
t(\mathbf{X} | \mathbf{Y})= i_{+1}(\mathbf{X} ; \mathbf{Y}), 
\end{equation*}
which is equivalent to (\ref{eq12}). 
\label{pro2}
\end{proposition}

Another interesting case is when both $(\mathbf{X},\mathbf{Y})$ and $\mathbf{Y}$ are Markov. 
In this case, a necessary and sufficient condition for (\ref{eq12}) can be derived easily: 

\begin{proposition}
If both $(\mathbf{X},\mathbf{Y})$ and $\mathbf{Y}$ are finite-alphabet stationary ergodic Markov processes, 
then we have 
\begin{equation*}
t(\mathbf{X} | \mathbf{Y})= i_{+1}(\mathbf{X} ; \mathbf{Y}) 
\end{equation*}
if and only if the condition 
\begin{equation}
p(x_2,y_2 | x_1,y_1)=p(x_2 | x_1,y_1)p(y_2 | y_1) 
\label{eq15}
\end{equation}
holds for any $x_1,x_2 \in A_n$ and $y_1,y_2 \in A_m$. 
\label{pro3}
\end{proposition}

\textit{Proof. }
Proving the claim is equivalent to showing 
\begin{equation*}
h(\mathbf{X} | \mathbf{Y})=h(\mathbf{X},\mathbf{Y})-h(\mathbf{Y}). 
\end{equation*}
By using the Markov property, we obtain 
\begin{align*}
& h(\mathbf{X} | \mathbf{Y}) - h(\mathbf{X},\mathbf{Y}) + h(\mathbf{Y}) \\
&= H(X_1,X_2 | Y_1) - H(X_1,X_2 | Y_1,Y_2) \geq 0. 
\end{align*}
In the last inequality, we have the equality if and only if 
$Y_2$ is independent of $X_1^2$ given $Y_1$, that is, 
\begin{equation}
p(x_1,x_2,y_2 | y_1)=p(x_1,x_2 | y_1)p(y_2 | y_1) 
\label{eq17}
\end{equation}
for any $x_1,x_2 \in A_n$ and $y_1,y_2 \in A_m$, which is equivalent to the condition in the proposition. 

\hfill $\Box$\\

Let us introduce the \textit{symbolic time-delayed mutual information rate} by 
\begin{equation}
i_{+1}^*(\mathbf{X} ; \mathbf{Y}) \equiv \lim_{L \to \infty} \frac{1}{L} I^*(X_1^{L+1};Y_1^L), 
\end{equation}
where 
\begin{equation*}
I^*(X_1^{L+1};Y_1^L) \equiv H^*(X_1^{L+1})+H^*(Y_1^L)-H^*(X_1^{L+1},Y_1^L) 
\end{equation*}
and discuss the relationship with the transfer entropy rate, the symbolic transfer entropy rate and 
the (time-delayed) mutual information rate. 
$i_{+1}^*(\mathbf{X} ; \mathbf{Y})$ 
exists for any bivariate finite-alphabet stationary stochastic process as we will see below. 

Similar properties with the time-delayed mutual information rate hold for the symbolic time-delayed 
mutual information rate: 
first, we note that $t^*(\mathbf{X} | \mathbf{Y}) \leq i_{+1}^*(\mathbf{X};\mathbf{Y})$ holds 
for any bivariate finite-alphabet stationary stochastic process $(\mathbf{X},\mathbf{Y})$ 
such that $t^*(\mathbf{X} | \mathbf{Y})$ exists because 
the similar inequality with (\ref{eq10-1}) holds for permutation analogues of corresponding quantities. 
Second, the symbolic time-delayed mutual information rate also admits the following expression: 
\begin{equation}
i_{+1}^*(\mathbf{X} ; \mathbf{Y}) = h^*(\mathbf{X}) + h^*(\mathbf{Y}) - h^*(\mathbf{X},\mathbf{Y}), 
\label{eq18}
\end{equation}
where $h^*(\mathbf{X},\mathbf{Y})= \lim_{L \to \infty} H(R_1^L,S_1^L)/L$. 
Thus, if we introduce the \textit{symbolic mutual information rate} between $\mathbf{X}$ and $\mathbf{Y}$ by 
\begin{equation*}
i^*(\mathbf{X} ; \mathbf{Y})=\lim_{L \to \infty} \frac{1}{L}I^*(X_1^L;Y_1^L), 
\end{equation*}
then we have 
\begin{equation*}
i_{+1}^*(\mathbf{X} ; \mathbf{Y}) = i_{+1}^*(\mathbf{Y} ; \mathbf{X}) = i^*(\mathbf{X} ; \mathbf{Y}) = i^*(\mathbf{Y} ; \mathbf{X}). 
\end{equation*}
Since the symbolic time-delayed mutual information rate is a sum of permutation entropy rates, 
we have 
\begin{equation*}
i_{+1}(\mathbf{X} ; \mathbf{Y}) = i_{+1}^*(\mathbf{X} ; \mathbf{Y}) 
\label{eq19}
\end{equation*}
for any bivariate finite-alphabet stationary stochastic process $(\mathbf{X},\mathbf{Y})$. 

By combining Theorem \ref{thm1}, Proposition \ref{pro2}, Proposition \ref{pro3} and (\ref{eq19}), 
we obtain the following proposition: 

\begin{proposition}
Let $(\mathbf{X},\mathbf{Y})$ be a bivariate finite-alphabet stationary ergodic Markov process 
over the alphabet $A_n \times A_m$. Then, 
\begin{itemize}
\item[(i)]
if 
\begin{equation*}
p(x_2,y_2 | x_1,y_1)=p(x_2 | x_1,y_1)p(y_2 | x_1,y_1) 
\end{equation*}
holds for any $x_1,x_2 \in A_n$ and $y_1,y_2 \in A_m$, 
then we have 
\begin{equation*}
t^*(\mathbf{X}|\mathbf{Y})=t(\mathbf{X}|\mathbf{Y})=i_{+1}(\mathbf{X} ; \mathbf{Y})=i_{+1}^*(\mathbf{X} ; \mathbf{Y}). 
\end{equation*}
\item[(ii)]
If $\mathbf{Y}$ is a Markov process and 
\begin{equation*}
p(x_2,y_2 | x_1,y_1)=p(x_2 | x_1,y_1)p(y_2 | y_1) 
\end{equation*}
holds for any $x_1,x_2 \in A_n$ and $y_1,y_2 \in A_m$, 
then we have 
\begin{equation*}
t^*(\mathbf{X}|\mathbf{Y})=t(\mathbf{X}|\mathbf{Y})=i_{+1}(\mathbf{X} ; \mathbf{Y})=i_{+1}^*(\mathbf{X} ; \mathbf{Y}). 
\end{equation*}
\end{itemize}
\label{pro4}
\end{proposition}

\section{\label{sec5}Concluding Remarks}
In this paper, we proved that the equality between the transfer entropy rate and 
the symbolic transfer entropy rate holds for any bivariate finite-alphabet stationary 
ergodic Markov process. The assumption is strong, however, it is the first theoretical result 
on permutation analogues of the transfer entropy rate. We also discussed the relationship between 
these quantities and the time-delayed mutual information rate and its permutation analogue. 

Next natural question is how we can weaken the assumption for (\ref{eq5}). 
At present, the authors are aware that the equality (\ref{eq5}) can be at least extended to 
any finite-state finite-alphabet hidden Markov process whose state transition matrix is 
irreducible by a similar, but technically improved discussion as in the ergodic Markov case. 
The crucial point is how to overcome $O(\log L)$ diverging term arising in the difference 
between the block Shannon entropy and its permutation analogue. 
Research results along this line will be presented elsewhere. 

Our result in this paper is restricted. However, we illustrated how the duality between $\phi_n$ and $\mu$, 
which is called the duality between values and orderings in \cite{Haruna2011}, can be used 
in different setting considered in \cite{Haruna2011}. We hope that the duality method 
opens up a systematic study on the relationship between the information-theoretic quantities 
and their permutation analogues for finite-state stationary stochastic processes. 

\section*{Acknowledgement}
TH was supported by JST PRESTO program. 
The authors would like to thank D. Kugiumtzis for his useful comments and discussion. 
The authors also acknowledge anonymous referees for their helpful comments to improve the manuscript.

\end{document}